\documentclass[12pt,preprint,aps,nofootinbib,showpacs]{revtex4}
\usepackage{graphics}
\usepackage{psfig}
\newcommand*{\be}{\begin{equation}}
\newcommand*{\ee}{\end{equation}}
\providecommand*{\ler}{\stackrel{\scriptstyle <}{\scriptstyle \sim}}

\begin{document}

\title{ Constraints On Radiative Neutrino Mass Models From Oscillation Data}

\author{ Probir Roy }
\email{probir@theory.tifr.res.in}
\author{Sudhir K. Vempati }
\email{sudhir@theory.tifr.res.in}
\affiliation{ Department of Theoretical Physics,\\
 Tata Institute of Fundamental Research,\\
Colaba, Mumbai 400 005, India. }

\preprint{TIFR-TH/01-46}
\begin{abstract}
The three neutrino Zee model and its extension
including three active and one sterile species are studied in the 
light of new neutrino oscillation data. We obtain analytical relations
 for the mixing angle in solar oscillations in terms of neutrino mass 
squared differences. For the four neutrino case, we obtain the result 
$\mathsf{ sin^2 2 \theta_\odot \approx
 1 - \left[ (\Delta m^2_{Atm})^2/(4~ \Delta m^2_{LSND} \Delta m^2_\odot) 
\right]^2 }$, which can accommodate both the large and small
mixing scenarios. We show that within this framework, while both the SMA-MSW 
and the LMA-MSW solutions can easily be accommodated, it would be difficult 
to reconcile the LOW-QVO solutions. We also comment on the active-sterile 
admixture within phenomenologically viable textures.
\end{abstract}


\maketitle

\section{ Introduction}

Recent results from SNO \cite{sno} and super-Kamiokande \cite{sksolar} 
experiments have confirmed the presence of a non-electron flavor in
the measured solar $\nu_e$ flux on earth implying the existence of neutrino
oscillations.  A detailed analysis of the data 
favors the Large Mixing Angle (LMA) solution \cite{dataanal} within the 
Mikheev-Smirnov-Wolfenstien (MSW) framework, though other solutions are
still not ruled out. A simplified two-flavor
oscillation picture prefers \cite{dataanal} 
a neutrino mass squared difference 
$\mathsf{\Delta m^2_\odot \sim 4 \times 10^{-5}~ eV^2}$ 
with $\mathsf{ sin^2 2 \theta_\odot
 \sim 0.66}$ for the mixing angle $\mathsf{\theta_\odot}$. On the other hand,
results \cite{skatmos} from the super-K atmospheric neutrino experiments
 indicate a non-muonic component in the atmospheric $\mathsf{\nu_\mu}$ flux.
 Again, in terms of the difference of squared masses between two oscillating
 neutrinos, these imply 
$\mathsf{\Delta m^2_{Atm} ~\sim ~3 ~\times 10^{-3}~ eV^2}$~
 with a near maximal mixing $\mathsf{sin^2 2 \theta_{Atm} ~\sim 1}$. These two
independent scales and mixing angles can be accommodated within the standard
model of three neutrinos $\mathsf{\nu_e, \nu_\mu, \nu_\tau}$. However,
in addition to  these, another scale pertaining to neutrinos is indicated
by $\mathsf{\bar{\nu_\mu} \leftrightarrow \bar{\nu_e}~ (and~ \nu_\mu 
\leftrightarrow \nu_e) }$ flavor oscillations seen by the LSND 
experiment \cite{lsndexp}. The parameters of a two-flavor oscillation 
that explain these results are 
$\mathsf{\Delta m^2_{LSND} \sim }~ \mathcal{O}(\mathsf{1~eV^2})$ and 
$\mathsf{sin^2 2 \theta_{LSND} \sim }~ \mathcal{O}(\mathsf{10^{-3}})$. A
simultaneous explanation of all the three types of oscillations is not possible
with only three types of  neutrinos. 

An additional light neutrino, which is not electroweak active and is hence 
called the sterile neutrino $\mathsf{\nu_s}$, is often proposed to understand
all the above anomalies simultaneously.  Within this four neutrino framework,
only two distinct mass patterns ~(the \textbf{3 + 1} and the \textbf{2 + 2}) 
were originally allowed by the data. 
 The former has three nearly degenerate active neutrinos, with mass
differences equal to the solar and atmospheric oscillation scales, separated 
as a whole in mass from the sterile flavor by the large LSND mass 
difference. However, it is not favored when fitted \cite{lsndanal} 
to the latest 
LSND results, together with data from reactor experiments like BUGEY; 
 therefore, we discard it. In the \textbf{2 + 2} case, two doublets, 
each nearly degenerate
with mass splits $ \mathcal{O}(\mathsf{\sqrt{\Delta m_\odot^2}})$ and 
$ \mathcal{O}(\mathsf{\sqrt{\Delta m_{Atm}^2}})$ and corresponding mixing angles
$\mathsf{\theta_\odot}$ and  $\mathsf{\theta_{Atm}}$ 
respectively,  are separated by a
large mass difference $ \mathcal{O}(\mathsf{\sqrt{\Delta m_{LSND}^2}})$.
The mixing between the two doublets is controlled by the angle 
$\mathsf{\theta_{LSND}}$. 

\begin{figure}[t]
\includegraphics{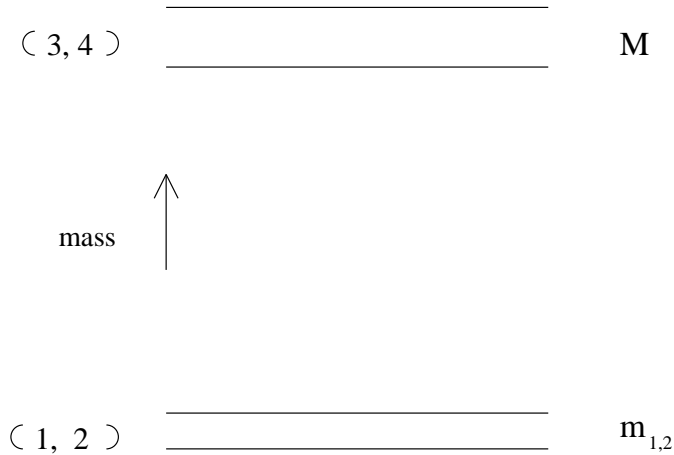}
\caption{The $\mathsf{2 + 2}$ neutrino mass pattern. The heavier 
(lighter) pair is split by the mass scale of the atmospheric (solar) 
neutrino anomaly while the two pairs are separated by the LSND scale.} 
\end{figure}

It is true that, strictly within two flavor oscillations, neither the SNO 
nor the super-K results support the occurrence of a sterile species in 
either solar
or atmospheric neutrino oscillation. 
Nevertheless, this simple picture changes when one considers all four neutrinos
$\mathsf{\nu_e,\nu_\mu,\nu_\tau,\nu_s}$.
A comprehensive analysis within a four neutrino framework of all (including
the latest) data on solar, atmospheric and LSND oscillations, also taking into 
account extant reactor and accelerator constraints, has recently been 
performed by 
Gonzalez-Garcia \textit{et.al} \cite{concha}. To summarize their results,
it is convenient to define the four species neutrino mixing matrix as

\be
\label{4gmns}
\mathsf{U} = 
\left( 
\begin{array}{cccc}
\mathsf{U_{e1}}&\mathsf{U_{e2}}&\mathsf{U_{e3}}&\mathsf{U_{e4}} \\
\mathsf{U_{\mu 1}}&\mathsf{U_{\mu 2}}&\mathsf{U_{\mu 3}}&\mathsf{U_{\mu4}} \\
\mathsf{U_{\tau 1}}&\mathsf{U_{\tau 2}}&\mathsf{U_{\tau 3}}&
\mathsf{U_{\tau 4}}\\
\mathsf{U_{s1}}&\mathsf{U_{s2}}&\mathsf{U_{s3}}&\mathsf{U_{s4}} 
\end{array}
\right). 
\ee 

\noindent
In eq.(\ref{4gmns}), the subscripts $\mathsf{(1,2)}$ represent the lighter
pair participating in solar neutrino oscillations, while $\mathsf{(3,4)}$ refer
to the heavier pair relevant to atmospheric neutrino oscillations.
These pairs are separated in mass by the LSND mass scale. We present a 
schematic diagram of this scenario in Fig.1.
For that situation, the BUGEY experiment \cite{bugey} provides
the maximum constraint on the $\mathsf{\nu_e}$ content in the 
heavier $\mathsf{(3,4)}$ pair namely

\be
\label{bugeyconst}
\mathsf{ |U_{e3}|^2 + |U_{e4}|^2 \ler 10^{-2} .}
\ee

\noindent
In addition, the CCFR \cite{ccfr} and the CDHSW \cite{cdhsw} experiments
constrain the $\mathsf{\nu_\mu }$ content in the lighter $\mathsf{(1,2)}$ 
pair by 

\be
\label{ccfrp}
\mathsf{ |U_{\mu 1}|^2 + |U_{\mu 2}|^2 \ler ~0.2 }, 
\ee

\noindent
again for an LSND scale mass separation between the pairs.

The rest of the constraints on $\mathsf{U}$ arrive from solar and 
atmospheric neutrino
oscillation data. Eqs. (\ref{bugeyconst}) and (\ref{ccfrp}), show that in
the \textbf{2 + 2} mass pattern, $\mathsf{\nu_e}$ is mostly confined 
to the $\mathsf{(1,2)}$
pair whereas $\mathsf{\nu_\mu}$ largely resides in the $\mathsf{(3,4)}$ pair.
Three alternative situations are now possible \cite{smirperes}: 

\vskip 0.4cm
\noindent
(i)~ Solar neutrino oscillations take place into the purely active 
neutrino $\mathsf{\nu_\tau}$; 
atmospheric neutrinos oscillate into the purely sterile neutrino 
$\mathsf{\nu_s}$ with maximal mixing. 

\vskip 0.4cm
\noindent
(ii)~ Solar neutrino oscillations take place into the purely sterile neutrino
$\mathsf{\nu_s}$;  atmospheric neutrinos get converted to the purely
active neutrino $\mathsf{\nu_\tau}$ with maximal mixing. 

\vskip 0.4cm
\noindent
(iii)~ Both solar and atmospheric neutrinos oscillate into linear combinations
of $\mathsf{\nu_s}$ and $\mathsf{\nu_\tau}$; the combination pertaining to
atmospheric neutrino oscillations is maximally mixed with $\mathsf{\nu_\mu}$. 

\vskip 0.4cm
As mentioned earlier,  recent data from both the SNO \cite{sno} and 
the super-K \cite{skatmos} experiments disfavor two flavor neutrino 
oscillations into only a sterile species as an explanation for either the solar
or the atmospheric anomaly. Consequently, configurations 
(i) and (ii) are severely constrained. In contrast, configuration (iii) 
can still be realized \cite{concha} even with new data from SNO \cite{sno}.
The favorite \textbf{2 + 2} configuration, most favored by the data, is when  
the linear combination of $\mathsf{\nu_s}$ and $\mathsf{\nu_\tau}$ 
in the state to which $\mathsf{\nu_e}$ from the Sun oscillates, is not
maximal but in the ratio $\mathsf{1:2}$, \textit{i.e,} $\mathsf{20:80}$
in the probabilities.  We shall define this `active-sterile admixture'
as the sterile content in the solar sector: 

\be
\label{actst}
\mathsf{
A \equiv  |U_{s1}|^2 + |U_{s2}|^2. 
}
\ee

\noindent
When the solar mixing angle is within the LMA region, corresponding to
the best-fit solution, the active-sterile admixture is required to be
\cite{concha}: 

\be
\label{admix}
\mathsf{
A \approx 0.18 - 0.2.
}
\ee 

\noindent
This means that the `atmospheric' pair $\mathsf{(3,4)}$ has a dominant
sterile content. To be precise, the sterile species participates in about
$\mathsf{ 20 \%~ (80 \%)  }$ of the solar~ (atmospheric) neutrino oscillations.
It should be emphasized that the SNO results have played a key role in 
these conclusions\footnote{ The active (sterile) content in the 
solar neutrino sector can be decreased (increased) \cite{barger} to more
than what is implied by eq.(\ref{admix}) if the $\mathsf{^8 B}$ flux is 
suitably renormalized. Such a renormalization of up to $\mathsf{ 30 \%}$ may
be allowed \cite{bahcall} by the theoretical 
uncertainties in the calculation of the 
said flux. It has been shown recently \cite{dataanal}, however, that 
the inclusion of the SK data on the day and night spectral energy distribution 
disfavor
such a scenario at $\mathsf{3 \sigma}$ level.} reached by the authors of 
Ref.\cite{concha}. This configuration leads to what has been  called 
\cite{concha} \textit{close to active solar
plus close to sterile atmospheric} ~(\textbf{CAS + CSA}) 
neutrino oscillations.

In addition to the above `best-fit' solution, configuration (ii) with 
the solar neutrino oscillations occurring into purely sterile neutrinos
is still allowed by the data. The active-sterile 
admixture $\mathsf{A}$ in this case is : 

\be
\mathsf{ A = |U_{s1}|^2 + |U_{s2}|^2 \approx 0.91 - 0.97 }.
\ee

\noindent
However, here the solar neutrino mixing angle lies in the SMA
region. For both of these fits, it is seen that  

\be
\label{bestfit}
\mathsf{ |U_{\mu 1}|^2 + |U_{\mu 2}|^2} \approx \mathsf{0.} \\
\ee

In this work, we focus on these mass patterns and study the viability of 
the $\mathsf{ 4 \times 4}$ neutrino mass matrix within a radiative
Zee-type model \cite{zee1} extended to include the sterile neutrino \cite{gaur}.
An extension from the standard $\mathsf{3 \times 3}$ mass matrix to the
$\mathsf{4 \times 4}$ mass matrix  can be 
realized either by a conventional seesaw type of mechanism \cite{mapr}
 with heavy right-handed states or in a radiative model \cite{gaur}. 
The three neutrino Zee model has been quite popular in analyzing neutrino 
oscillation data (minus those of the LSND experiment) on account of its 
predictivity.  However, this model has run into a serious problem with the
data on the mixing angle pertinent to solar neutrino oscillations. It has
been shown \cite{jarlskog} that $\mathsf{sin^2 2 \theta_\odot }$ in this
model is forced to be close to unity \cite{koide} within 
$\mathcal{O} [(\mathsf{\Delta m^2_\odot / \Delta m^2_A})^2]$ in disagreement
with the best-fit value \cite{dataanal} $\mathsf{sin^2 2 \theta_\odot  \sim
0.66 }$. More specifically, if $\mathsf{sin^2 2 \theta_{Atm} \sim 1 }$ is
used as an input, the Zee model is found to allow \cite{jarlskog,koide} only
solutions with bimaximal mixing \cite{bimax}. 
This result of the Zee model is a natural consequence of the 
structure of its mass matrix - specifically, its
vanishing diagonal elements, rather than the details of the model. It is
thus essential to understand the consequences of such a mass matrix when 
extended to the four neutrino case. To be precise, one needs to compute
the allowed values of the mixing angle relevant to the solar sector
the four neutrino radiative model.

Within the standard three neutrino set up, various extensions of the Zee 
model have been presented in literature
to evade the compulsion of $\mathsf{sin^2 2 \theta_\odot \approx 1}$. Most
of these include additional couplings and/or fields \cite{zeeext} and then 
they can avoid bimaximal mixing and incorporate the LMA solution. 
Here we consider the extension of the Zee model by an additional sterile
species \cite{gaur}. This would enable us to include solutions corresponding
to the LSND anomaly. Using the property of vanishing diagonal elements,
we are able to make a prediction on the solar neutrino mixing angle involving
all the three squared mass differences, which is compatible with the present
data.  We show that the model is suitable
for explaining all the three neutrino anomalies. The precise value of 
$\mathsf{\theta_\odot}$ depends on the ratio between the square of the 
atmospheric  neutrino mass squared difference and the product of the solar
and LSND mass squared differences. Present experimental errors on these
measured mass squared differences allow the solar neutrino
oscillations to take place with MSW conversion either with small or 
large mixing.  However, we find it difficult to reconcile the LOW-QVO 
solutions with this scenario, though they might not be completely ruled out. 

The rest of the paper is organized as  follows. We show in section \textbf{II}
how the near-maximality constraint on $\theta_\odot$ arises from the vanishing
diagonal elements of the $\mathsf{3 \times 3}$ Zee mass matrix, 
establishing a method that can be 
extended to the $\mathsf{4 \times 4}$ case. This extension is done in 
section \textbf{III} where we derive the result for 
$\mathsf{sin^2 2\theta_\odot}$
mentioned in the abstract. In section \textbf{IV}, we comment on
textures of the mass matrix within this model which are phenomenologically
viable. We then show that the active-sterile admixture in solar neutrinos 
and the LSND mass scale within these models have a common origin within
the mass matrix.  The final section, section $\mathsf{V}$ summarizes our 
conclusions.

\section{Restriction on solar neutrino mixing from the 
three neutrino Zee Mass matrix} 

As already mentioned, the three neutrino Zee model has a serious
problem on account of the current neutrino oscillation data, specifically with
the data from the solar sector. The reason for this can be 
seen, without going in to the details of the model, from one feature of it - 
namely, its vanishing diagonal elements.
The Zee mass matrix in three generations is given, in the
$\mathsf{ \nu_e, \nu_\mu, \nu_\tau}$ flavor basis, by 

\be
\label{zee3gmass}
\mathcal{M}_{\mathsf{\nu}}^{\mathsf{(3)}} =  
\left( 
\begin{array}{ccc}
\mathsf{0}&\mathsf{a}&\mathsf{b} \\
\mathsf{a}&\mathsf{0}&\mathsf{c} \\
\mathsf{b}&\mathsf{c}&\mathsf{0}
\end{array}
\right). 
\ee 

\noindent
The elements $\mathsf{a,~b,~c}$ are three\footnote{For three neutrinos,
the  Zee mass matrix has only three parameters. Thus all  phases 
can be rotated away from the mixing matrix, by absorbing them in the
three neutrino fields. Hence one can choose the orthogonal matrix
 $\mathsf{O_{MNS}}$ for mixing instead of the
unitary  Maki-Nakagawa-Sakata matrix $\mathsf{U_{MNS}}$.} 
real parameters determining two mass squared differences and 
three mixing angles relevant to solar and
atmospheric neutrino oscillation phenomenology. 
Evidently, this leads to a constrained pattern of neutrino
masses and mixing which has to face the challenge of the emerging data.
Three relations emerge from the vanishing diagonal terms\footnote{
Two loop corrections can induce very small nonzero diagonal entries 
\cite{chang} in the Zee mass matrix of eq. (\ref{zee3gmass}). Strictly
speaking, they would change eqs. (\ref{3gconst1}), (\ref{3gconst3} and 
(\ref{3gconst2}). However estimates \cite{chang} of these tiny
changes suggest that they would not significantly affect the statement
of eq.(\ref{bound}) on the solar neutrino mixing angle 
$\mathsf{\theta_\odot}$ \cite{koide}. }
of $\mathcal{M}_\mathsf{\nu}^\mathsf{(3)}$ if we write 
$\mathcal{M}_\mathsf{\nu}^\mathsf{(3)} 
= \mathsf{O_{MNS}~diag \{ m_{\nu_1},
m_{\nu_2}, m_{\nu_3} \}~O^T_{MNS}}$ :

\begin{eqnarray}
\label{unit1a}
\mathsf{m_{\nu_1} O_{e 1}^2 + m_{\nu_2} O_{e 2}^2 + 
m_{\nu_3} O_{e 3}^2}&=& \mathsf{0},\\
\label{unit1b}
\mathsf{m_{\nu_1} O_{\mu 1}^2 + m_{\nu_2} O_{\mu  2}^2 + m_{\nu_3} 
O_{\mu 3}^2}&=& \mathsf{0},\\
\label{unit1c}
\mathsf{m_{\nu_1} O_{\tau 1}^2 + m_{\nu_2} O_{\tau  2}^2 
+ m_{\nu_3} O_{\tau 3}^2}&=& \mathsf{0}.
\end{eqnarray}

Because of the orthogonality of $\mathsf{O_{MNS}}$, eqs. (\ref{unit1a}),
(\ref{unit1b}), (\ref{unit1c}) lead to \cite{jarlskog}:

\begin{eqnarray}
\label{3gconst1}
\mathsf{m_{\nu_1} + m_{\nu_2} + m_{\nu_3}}&=&\mathsf{0}, \\[8 pt]
\label{3gconst3}
\mathsf{ \left( O_{e1}^2 - O_{e3}^2 \right) ~
\left( O_{\mu 2}^2 - O_{\mu 3}^2 \right) } &=& \mathsf{ \left( O_{\mu 1}^2 - 
O_{\mu 3}^2 \right) \left( O_{e2}^2 - O_{e3}^2 \right) }~, \\[8pt]
\label{3gconst2}
\mathsf{ {O_{e2}^2 - O_{e3}^2 \over O_{e1}^2 - O_{e3}^2 } } &= &
- \mathsf{{m_{\nu_1} \over m_{\nu_2}}}~ .
\end{eqnarray}

\noindent
Eq.(\ref{3gconst1}) is a trivial consequence of the tracelessness of 
$\mathcal{M}_{\mathsf{\nu}}^\mathsf{(3)}$ while eq.(\ref{3gconst3}) implies
that one of the mixing angles is not independent. If 
$\mathsf{ \Delta m^2_{ij} \equiv~| m_{\nu_i}^2 - m_{\nu_j}^2 |}$ and is 
positive by definition, let us choose 
$\mathsf{\Delta m_{32}^2 ~(= ~\Delta m^2_{Atm})}$ and 
$\mathsf{\Delta m_{21}^2 ~(= ~\Delta m^2_\odot)}$ as the two independent
mass squared differences. We know from condition (\ref{3gconst1}) that one 
the eigenvalues of $\mathcal{M}_\mathsf{\nu}^\mathsf{(3)}$ is not independent.
The latter is chosen to be $ \mathsf{m_{\nu_3}}$ and can be eliminated. 
Thus one can now write 

\begin{eqnarray}
\label{newconst3g1}
\mathsf{ | m_{\nu_1}^2 + 2 m_{\nu_1} m_{\nu_2} |}&=& 
\mathsf{\Delta m^2_{Atm}}~, \\
\label{newconst3g2}
\mathsf{| m_{\nu_2}^2 - m_{\nu_1}^2 | }&=&\mathsf{\Delta m^2_\odot}~.
\end{eqnarray}

\noindent
We emphasize here that the quantities 
appearing in the RHS of eqs. (\ref{newconst3g1}) and (\ref{newconst3g2}), 
namely $\mathsf{\Delta m^2_\odot, ~\Delta m^2_{Atm}}$ 
are  positive. These equations  yield $\mathsf{m_{\nu_1}}$ and
$\mathsf{m_{\nu_2}}$ in terms of $\mathsf{\Delta m^2_{Atm}}$ and 
$\mathsf{\Delta m^2_\odot}$. Let us define 

\be
\mathsf{
\alpha \equiv {\Delta m^2_\odot \over \Delta m^2_{Atm} }~. 
}
\ee 

\noindent
In order to determine the solar mixing angle, we can take 
$\mathsf{\theta_{23}~ =~  \theta_{Atm}}$, \textit{i.e} the atmospheric 
mixing angle, to be exactly maximal. Furthermore, on account of the 
CHOOZ \cite{chooz} experimental result, we can put 
$\mathsf{O_{e3} = \epsilon \ler 0.1}$, keeping only 
$\mathcal{O}(\epsilon)$ terms in $\mathsf{O_{MNS}}$. 
Choosing the  mixing angle $\mathsf{\theta_{12} = \theta_\odot}$ for solar 
neutrino oscillations, $\mathsf{O_{MNS}}$ now gets reduced to

\be
\label{3gmns}
\mathsf{O_{MNS}} =  
\left( 
\begin{array}{ccc}
\mathsf{cos~ \theta_\odot}&\mathsf{sin~ \theta_\odot}&\mathsf{~~\epsilon} \\
\mathsf{(-\epsilon~ cos~ \theta_\odot - sin~ \theta_\odot )/\sqrt{2}}&
\mathsf{(-\epsilon~ sin~ \theta_\odot + cos~\theta_\odot )/\sqrt{2}}&
\mathsf{~~1/\sqrt{2}} \\
\mathsf{(-\epsilon~ cos~ \theta_\odot + sin~\theta_\odot )/\sqrt{2}}&
\mathsf{(-\epsilon~ sin~ \theta_\odot -cos~ \theta_\odot )/\sqrt{2}}&
\mathsf{~~1/\sqrt{2}}
\end{array}
\right) + \mathcal{O}(\mathsf{\epsilon^2 })~.
\ee 

\noindent
The substitution of  eq.(\ref{3gmns}) into eq.(\ref{3gconst2}) leads after 
some algebra to

\be
\label{solang}
\mathsf{
sin^2 2 \theta_\odot = - { {4~m_{\nu_1}/m_{\nu_2}} \over
(1 - {m_{\nu_1}/ m_{\nu_2}})^2 
}} + \mathcal{O} \left(\mathsf{\epsilon^2 ( 1 + m_{\nu_1} 
/m_{\nu_2} ) } \right)~. 
\ee 

\noindent
The physical solutions\footnote{
We choose only those solutions which yield positive values for 
$\mathsf{m_{\nu_1}^2}$, $\mathsf{m_{\nu_2}^2}$ and 
$\mathsf{ sin^2 2 \theta_\odot}$.  There are four sets with
$\mathsf{ m_{\nu_1} = \pm (\Delta m^2_{Atm}/3)^{1 \over 2}
~ ( 1 + 2 \alpha + 2 \sqrt{ 1 + \alpha + \alpha^2 })}$ and 
$\mathsf{ m_{\nu_2} = - ( - \alpha + \sqrt{ 1 + \alpha + \alpha^2 })~
m_{\nu_1}}$, 
$\mathsf{ m_{\nu_1} = \pm (\Delta m^2_{Atm}/3)^{1 \over 2}
~ ( 1 - 2 \alpha + 2 \sqrt{ 1 - \alpha + \alpha^2 })}$ and 
$\mathsf{ m_{\nu_2} = - ( \alpha + \sqrt{ 1 - \alpha + \alpha^2 })~
m_{\nu_1}}$, 
expressed as functions of $\mathsf{\alpha}$ 
while  $\mathsf{m_{\nu_3}}$ is always $\mathsf{- m_{\nu_1} -
m_{\nu_2}}$. The solutions 
are physical only for 
$\mathsf{|m_{\nu_1}| \sim |m_{\nu_2}| \gg |m_{\nu_3}|}$
so that $\mathsf{\sqrt{\Delta m^2_{Atm}}}$ $\sim$ $\mathsf{|m_{\nu_{1,2}} -
m_{\nu_3}}|$ and $\mathsf{\sqrt{\Delta m^2_\odot}}$ $\sim$ 
$|\mathsf{m_{\nu_1} - m_{\nu_2}}|$. It is not difficult to see that this
implies $\mathsf{a}$ $\sim$ $\mathsf{b}$ $\approx$ $\mathcal{O}(\mathsf{\sqrt{
\Delta m^2_{Atm}}})$ and $\mathsf{c}$ $\approx$ $\mathcal{O}(\mathsf{
\Delta m^2_\odot/\sqrt{\Delta m^2_{Atm}}})$. 
} of  
 $\mathsf{m_{\nu_1},~m_{\nu_2}}$ of eqs. (\ref{newconst3g1}) and 
(\ref{newconst3g2}) yield 

\be
\label{mratio}
\mathsf{
- {m_{\nu_1} \over m_{\nu_2}} = { 1 \over \alpha + 
\sqrt{ 1 - \alpha + \alpha^2 } }~.
} 
\ee 

\noindent
Eqs.~(\ref{solang}) and (\ref{mratio}) lead to the following constraint on
the solar mixing angle\footnote{ We should also mention that eq.(\ref{bound})
was initially derived in Ref.\cite{lavoura} within a seesaw model which
has a mass texture different from that of the Zee model. 
We thank K. R. Balaji for pointing this out.}

\be
\label{bound}
\mathsf{
sin^2 2 \theta_\odot = 1 - {1 \over 16} \alpha^2 + }~~
\mathcal{O} \left( \mathsf{ \epsilon^2 \alpha }\right),
\ee 

\noindent
matching with the result of Ref. \cite{koide}. 
Since \cite{dataanal, skatmos} $\mathsf{ \alpha \equiv 
(\Delta m_\odot^2/\Delta m^2_{Atm} ) \ler 6 \times 10^{-1}}$ and 
$\mathsf{ \epsilon \ler 10^{-1} }$, 
we see that $\mathsf{sin^2 2 \theta_\odot}$ is forced 
to be very close to unity (\textit{i.e} 
$\mathsf{| m_{\nu_1}| \approx |m_{\nu_2}|}$ )~-
a situation disfavored \cite{dataanal} 
by the data\footnote{ Some detailed numerical analyses of the Zee model,
confronting the present data, have argued that the possibility 
$\mathsf{sin^2 2 \theta_\odot \approx 1 }$ is not completely ruled out
\cite{frampton,biswa}, if one insists on $\mathsf{ 99 \%}$ C.L. limits. },
though not completely ruled out. 
A point to note is that eq.(\ref{bound}) follows from eq.(\ref{3gmns}) 
by use of only the elements in the first row of $\mathsf{O_{MNS}}$. This
derivation does not use the fact that 
$\mathsf{\theta_{23}~ \equiv ~\theta_{Atm}}$ is $\approx \mathsf{\pi/4}$. 
It is, of course, true that the extra input of the latter immediately forces 
near bimaximal mixing in the Zee model, as mentioned in the Introduction. 
Another point to keep in mind is that the texture $\mathsf{|c|~ \ll ~|a| ~\sim
|b|}$, forced by the neutrino mass solutions given in footnote 4,  implies
an approximate global $\mathsf{L_e - L_\mu - L_\tau}$ symmetry \cite{jarlskog}.

\section{Solar neutrino mixing in the Radiative Mass Model
with four neutrinos}

The Majorana mass matrix with three active and one sterile species
can be generally represented in the 
$\mathsf{ \{ \nu_e,~\nu_\mu,~\nu_\tau,~\nu_s \}}$ flavor basis as

\be
\label{fullmat}
\mathcal{M}_{\mathsf{\nu}}^\mathsf{(4)} =  
\left( 
\begin{array}{cccc}
\mathsf{0}&\mathsf{a}&\mathsf{b}&\mathsf{d} \\
\mathsf{a}&\mathsf{0}&\mathsf{c}&\mathsf{e} \\
\mathsf{b}&\mathsf{c}&\mathsf{0}&\mathsf{f}\\
\mathsf{d}&\mathsf{e}&\mathsf{f}&\mathsf{0} 
\end{array}
\right). 
\ee 

%
%
 
\noindent
The elements $\mathsf{a,~b,~c,~d,~e,~f}$ appearing in eq.(\ref{fullmat})
get generated radiatively at one-loop and are given by 

\begin{eqnarray}
\label{abcdef}
\mathsf{a}&=& \mathsf{f_{e \mu} (m_\mu^2 - m_e^2 ) 
\left( { \mu v_2 \over v_1} \right) F(m_{\chi_1}^2, m_{\phi_1}^2)}, \\
\mathsf{b}&=&\mathsf{ f_{e \tau} (m_\tau^2 - m_e^2 ) \left( { \mu v_2 \over v_1}
\right) F(m_{\chi_1}^2, m_{\phi_1}^2)}, \\
\mathsf{c}&=& \mathsf{ f_{\mu \tau} (m_\tau^2 - m_\mu^2 ) \left( 
{ \mu v_2 \over v_1} \right) F(m_{\chi_1}^2, m_{\phi_1}^2)}, \\ 
\mathsf{d}&=&\mathsf{ \left( f_{e \tau} f'_{\tau} m_\tau + f_{e \mu} 
f'_\mu m_\mu \right) \mu' u~ F(m_{\chi_1}^2, m_{\chi_2}^2)}, \\ 
\mathsf{e}&=&\mathsf{ \left( f_{\mu \tau} f'_{\tau} m_\tau + f_{\mu e} f'_e 
 m_e \right) \mu' u ~F(m_{\chi_1}^2, m_{\chi_2}^2)}, \\ 
\mathsf{f} &=& \mathsf{ \left( f_{\tau \mu} f'_{\mu} m_\mu + f_{\tau e}
 f'_e  m_e \right) \mu' u~ F(m_{\chi_1}^2, m_{\chi_2}^2)},
\end{eqnarray}

\noindent
where $\mathsf{f,~ f'~ (\mu,~ \mu')}$ are dimensionless (dimensional)
couplings in the model, $\mathsf{v_{1,2}}$ are two Higgs VEVs and 
$\mathsf{m_{\chi_{1,2}}, m_{\phi_1}}$ are the three Higgs scalar masses,
as explained in Ref.\cite{gaur}. Moreover, the function 
$\mathsf{F(m_1,m_2)}$ is defined as

\begin{equation}
\mathsf{
F(m_1,m_2) = {1 \over 16 \pi^2} {1 \over m_1^2 - m_2^2 } \ln \left(
{m_1^2 \over m_2^2 } \right).
} 
\end{equation}
\noindent
For simplicity, we assume the matrix elements of eq. (\ref{fullmat}) to 
be real and ignore the presence of any CP-violation in the neutrino sector.

In the same manner, as shown in the previous section, one can 
arrive\footnote{As in the three neutrino radiative Zee model,
we expect that the effect of the two loop corrections to the four neutrino
radiative model mass matrix would be small and thus of not much consequence
to the results presented hereafter.}  at
the following relations by equating  $\mathcal{M}_{\mathsf{\nu}}^\mathsf{(4)}$
 with $\mathsf{ U~diag\{ m_{\nu_1},~m_{\nu_2},~m_{\nu_3},~m_{\nu_4} \}~U^T}$:

\begin{eqnarray}
\label{unit2}
\mathsf{m_{\nu_1} U_{e 1}^2 + m_{\nu_2} U_{e 2}^2 + 
m_{\nu_3} U_{e 3}^2 + m_{\nu_4} U_{e 4}^2 }&=& \mathsf{0},\\
\mathsf{m_{\nu_1} U_{\mu 1}^2 + m_{\nu_2} U_{\mu  2}^2 + m_{\nu_3} 
U_{\mu 3}^2 + m_{\nu_4} U_{\mu 4}^2 }&=& \mathsf{0},\\
\mathsf{m_{\nu_1} U_{\tau 1}^2 + m_{\nu_2} U_{\tau  2}^2 
+ m_{\nu_3} U_{\tau 3}^2 + m_{\nu 4} U_{\tau 4}^2 }&=& \mathsf{0}, \\
\label{unit2a}
\mathsf{m_{\nu_1} U_{s 1}^2 + m_{\nu_2} U_{s 2}^2 
+ m_{\nu_3} U_{s 3}^2 + m_{\nu 4} U_{s 4}^2 }&=& \mathsf{0}, 
\end{eqnarray}

\noindent
where $\mathsf{U_{ij}}$ are the elements of the
matrix $\mathsf{U}$ defined in eq.(\ref{4gmns}). Since we ignore the 
phases present in the elements of $\mathsf{U}$, we can utilize the
unitarity of $\mathsf{U}$ to obtain the following relations from 
eq. (\ref{unit2}) - (\ref{unit2a}).

\begin{equation}
\label{4gconst1}
\mathsf{m_{\nu_1} + m_{\nu_2} + m_{\nu_3} + m_{\nu_4}} =  \mathsf{0}, 
\end{equation}
\begin{eqnarray}
\mathsf{m_{\nu_2}}&=& -\mathsf{ { (U_{e1}^2 - U_{e4}^2 ) ~(U_{\mu 3}^2 - 
U_{\mu 4}^2 ) - (U_{e3}^2 - U_{e4}^2 )~( U_{\mu1}^2 - U_{\mu4}^2) \over
(U_{e2}^2 - U_{e4}^2)~(U_{\mu3}^2 - U_{\mu 4}^2) - (U_{e3}^2 - U_{e4}^2)
~(U_{\mu2}^2 - U_{\mu4}^2) }~ m_{\nu_1}},  \\ [8pt]
\mathsf{m_{\nu_3}}&=& - \mathsf{m_{\nu_1}~ { U_{e1}^2 - U_{e4}^2 \over
U_{e3}^2 - U_{e4}^2 }~-~m_{\nu_2}~{U_{e2}^2 - U_{e4}^2 \over U_{e3}^2
- U_{e4}^2 } } ,
\end{eqnarray}
\begin{eqnarray}
\label{4gconst2}
\mathsf{ ~~{ (U_{e2}^2 - U_{e4}^2)(U_{\mu 3}^2 - U_{\mu 4}^2) 
 - (U_{e3}^2 - U_{e4}^2)~(U_{\mu 2}^2 - U_{\mu 4}^2) \over 
 (U_{e1}^2 - U_{e4}^2)~(U_{\mu 3}^2 - U_{\mu 4}^2) - 
(U_{e3}^2 - U_{e4}^2) (U_{\mu 1}^2 - U_{\mu 4}^2) } } \nonumber \\
 =~{\mathsf{ (U_{\tau 2}^2 - U_{\tau 4}^2)( U_{e3}^2 - U_{e4}^2)
- (U_{e2}^2 - U_{e4}^2)(U_{\tau 3}^2 - U_{\tau 4}^2 )  \over
(U_{\tau 1}^2 - U_{\tau 4}^2)(U_{e3}^2 - U_{e4}^2) - 
(U_{e1}^2 - U_{e4}^2) (U_{\tau 3}^2 - U_{\tau 4}^2 ) } }.
\end{eqnarray}

\noindent
These are the four neutrino versions of eqs. (\ref{3gconst1}), 
(\ref{3gconst3}) and (\ref{3gconst2}).

Eq. (\ref{4gconst1}) shows us that the $\mathsf{2 + 2}$ mass pattern depicted in Fig.1 can be easily made compatible with this model if
$\mathsf{|m_{\nu_1}| \sim |m_{\nu_2}| \ll |m_{\nu_3}| \sim |m_{\nu_4}|}$ or
if $\mathsf{|m_{\nu_1}| \ll |m_{\nu_2}| \ll |m_{\nu_3}| \sim |m_{\nu_4}|}$. 
In either case, there are three independent mass squared differences which
we choose to be $\mathsf{\Delta m^2_{21} \equiv \Delta m^2_\odot,~
\Delta m^2_{43} \equiv \Delta m^2_{Atm}}$ and $\mathsf{\Delta m_{32}^2
\equiv \Delta m^2_{LSND} }$. 
We can now perform an extended version of the same eigenvalue analysis as
done in Section \textbf{II} for the three neutrino case. The three independent
eigenvalues are chosen here to be $\mathsf{m_{\nu_1}, m_{\nu_2}, m_{\nu_3}}$.
Eqs.~(\ref{newconst3g1}) and (\ref{newconst3g2}) now extend to

\begin{eqnarray}
\label{4gsimul1}
\mathsf{
| (m_{\nu_1} + m_{\nu_2})^2 + 2 m_{\nu_3} (m_{\nu_1} + m_{\nu_2})|} &=&
\mathsf{ \Delta m^2_{Atm}}, \\ 
\label{4gsimul2}
\mathsf{ | m_{\nu_3}^2 - m_{\nu_1}^2| }&=& \mathsf{\Delta m^2_{LSND}},\\ 
\label{4gsimul3}
\mathsf{ | m_{\nu_2}^2 - m_{\nu_1}^2 |}&=& \mathsf{ \Delta m^2_{\odot}},
\end{eqnarray}

\noindent 
where the RHS quantities of the above equations, namely
$\mathsf{\Delta m^2_{LSND}}$, $\mathsf{\Delta m^2_\odot}$, 
$\mathsf{\Delta m^2_{Atm} }$, are positive as in eqs. (\ref{newconst3g1}) 
and (\ref{newconst3g2}).  The physical mass eigenvalue solutions to the above 
three equations form four sets as detailed in the Appendix. From these
we see that

\begin{eqnarray}
\label{4gratiodel}
\mathsf{
- { m_{\nu 1} \over m_{\nu_2}}} &=&
\mathsf{ - {N_1^\pm \over N_2^\pm }} \nonumber \\[12pt]
\label{4gmatratio}
&\approx& \mathsf{{4~\Delta m^2_\odot~ \Delta m^2_{LSND} \mp 
(\Delta m^2_{Atm} )^2
\over 4~ \Delta m^2_{LSND}~ \Delta m^2_\odot \pm (\Delta m^2_{Atm})^2 }
+ \mathcal{O} ( \alpha ) ,}
\end{eqnarray}
where $\mathsf{N_{1,2}^\pm}$ are defined in the Appendix and 
the upper (lower) sign corresponds to sets 1, 2 (3, 4).

In order to arrive at the prediction on the solar neutrino 
mixing angle in the four neutrino scenario, we can choose\footnote{The 
matrix $\mathsf{U}$ has six independent angles and two
independent phases in general. But one has the freedom to choose a
parameterization in which the phases do not appear in $\mathsf{U_{e1}}$ and
$\mathsf{U_{e2}}$. In any event, we are ignoring the phases. }
a parameterization of the $\mathsf{4 \times 4}$ generalization of
 $\mathsf{U_{MNS}}$ matrix as \cite{yasuda} 

\be
\label{4gmns2}
\mathsf{U} = 
\left( 
\begin{array}{cccc}
\mathsf{cos \theta_\odot}&\mathsf{sin \theta_\odot}
&\mathsf{\epsilon}&\mathsf{\epsilon'} \\
\mathsf{U_{\mu 1}}&\mathsf{U_{\mu 2}}&\mathsf{U_{\mu 3}}&\mathsf{U_{\mu4}} \\
\mathsf{U_{\tau 1}}&\mathsf{U_{\tau 2}}&\mathsf{U_{\tau 3}}&
\mathsf{U_{\tau 4}}\\
\mathsf{U_{s1}}&\mathsf{U_{s2}}&\mathsf{U_{s3}}&\mathsf{U_{s4}} 
\end{array}
\right)~ 
+ \mathcal{O}(\epsilon^2, \epsilon^{'~2}, \epsilon \epsilon').
\ee

\noindent
In eq.(\ref{4gmns2}), the solar neutrino oscillations have been chosen to 
be between the physical states $\mathsf{ |\nu_1 >}$ and $\mathsf{|\nu_2 >}$ 
where the BUGEY constraint eq.(\ref{bugeyconst}) has been taken care of by
choosing $\mathsf{U_{e3} = \epsilon}$ and $\mathsf{U_{e4} = \epsilon'}$,
both\footnote{It should be noted that in the analysis presented in 
\cite{concha} both $\mathsf{U_{e3}}$ and $\mathsf{U_{e4}}$ are taken to 
be zero.} being $\mathsf{\ler}~ \mathcal{O}(10^{-2})$. We keep 
only $\mathcal{O}(\mathsf{\epsilon~, \epsilon'} )$ terms 
in the above mixing matrix.  The rest of the parameters, while otherwise 
arbitrary, will be required to satisfy unitarity and other experimental 
constraints later. 

One can now determine the mixing angle in the solar sector
in much the same manner as in the three neutrino case. Eq.(\ref{solang}) is
still valid to $\mathcal{O}(\mathsf{\epsilon^2, \epsilon^{'~2}})$ and can
be rewritten as\footnote{Since $\mathsf{ m_{\nu_1}/m_{\nu_2} = 
tan^2 \theta_\odot} + \mathcal{O}(\mathsf{\epsilon^2,~\epsilon^{'~2}})$, it 
follows that the RHS of eq.(\ref{4gratiodel}) must be real and positive
modulo $\mathsf{\epsilon^2,~\epsilon^{'~2}}$ terms.}

\be
\label{4gsolang}
\mathsf{
sin^2~2 \theta_\odot = {- 4 m_{\nu_1}/m_{\nu_2} \over 
(1 - m_{\nu_1}/m_{\nu_2})^2 } + \mathcal{O}(\epsilon^2,~ \epsilon^{'~2})
}
\ee

\noindent
Eqs. (\ref{4gmatratio}) and (\ref{4gsolang}) together yield:

\begin{figure}[htb]
\label{solsfig}
\centerline{\psfig{figure=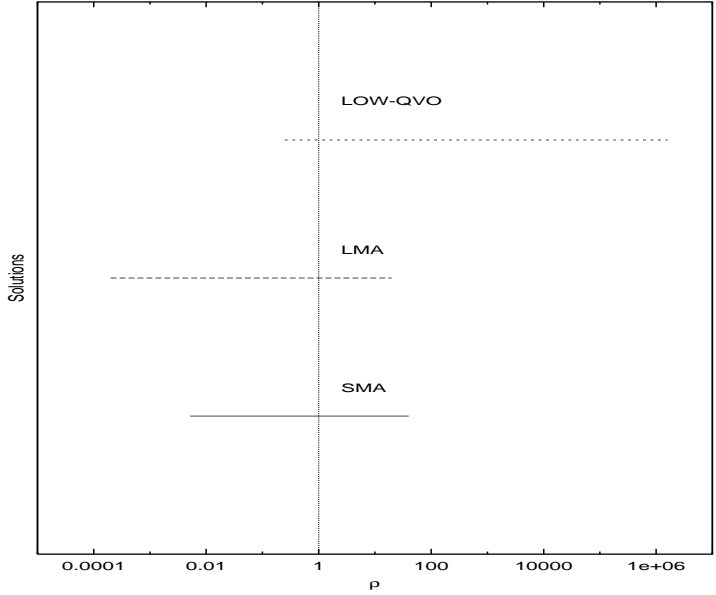,height=8 cm,width=10 cm}}
\caption{The range of $ \mathsf{ \rho \equiv {(\Delta m^2_{Atm})^2 \over ~4~ 
\Delta m^2_\odot ~ \Delta m^2_{LSND} ~} }$, shown on a logarithmic scale,
for different solar neutrino solutions.}
\end{figure}

\begin{equation}
\label{solpred}
\mathsf{
sin^2 2 \theta_\odot
\approx   1 - \rho^2 }
+ \mathcal{O} \mathsf{ \left( \alpha,~ 
\epsilon^2, ~\epsilon^{'~2} \right) }, 
\end{equation}

\noindent 
where $\mathsf{\rho}$ is defined as
\begin{equation}
\label{rho}
\mathsf{
\rho \equiv {(\Delta m^2_{Atm})^2 \over ~4~ 
\Delta m^2_\odot ~ \Delta m^2_{LSND} ~} }. 
\end{equation}

\noindent
Eq.(\ref{solpred}) leads us to the following conclusions.

\begin{itemize}
\item  Because of eq.(\ref{solpred}), we have the condition
\begin{equation}
\label{mainconst}
\mathsf{
\rho \equiv {(\Delta m^2_{Atm})^2 \over ~4~ 
\Delta m^2_\odot ~ \Delta m^2_{LSND} ~} \ler 1,
}
\end{equation}
upto $\mathsf{\alpha, \epsilon^2, \epsilon^{'~2}}$ terms.
In Fig.2 we plot the ranges of $\mathsf{\rho}$ for various proposed solutions
of solar neutrino oscillations. We have chosen $\mathsf{\Delta m^2_{LSND}
~\approx ~(0.2 - 2) eV^2}$ \cite{lsndexp}, $\mathsf{\Delta m^2_{Atm} ~\approx
~(1 - 8) \times 10^{-3} eV^2}$ \cite{skatmos} and, for the various solar 
neutrino solutions, 
$\mathsf{\Delta m^2_\odot (SMA) \approx (2 \times 10^{-6} - 2
 \times 10^{-5}) ~eV^2} $, 
$\mathsf{\Delta m^2_\odot (LMA) \approx (4 \times 10^{-6} - 6
 \times 10^{-4}) ~eV^2} $ 
$\mathsf{\Delta m^2_\odot ~(LOW-QVO) ~\approx ~(5 \times 10^{-7} - 5
 \times 10^{-11}) ~eV^2} $ ~\cite{dataanal}. From the figure, we see that
eq.(\ref{mainconst}) is clearly valid for large domains of 
SMA and LMA solutions. However, in the case of
LOW-QVO solutions, only a small domain obeys 
eq.(\ref{mainconst}) making it difficult to be accommodated within the 
$\mathsf{4 \times 4}$ radiative \cite{gaur} model of neutrino masses. 
A detailed numerical analysis of all oscillation data within the 
four neutrino model would be able to 
show the confidence level at which the LOW-QVO solution is valid given 
the mass matrix of eq. (\ref{fullmat}).

\item  As in section \textbf{III}, we have expressed the solar neutrino
mixing angle, $\mathsf{\theta_\odot}$ 
in terms of ratios of mass squared differences, though the
form is different.  However, \textit{ unlike in the three generation case, the 
solar neutrino mixing is not forced to be very nearly maximal.} 
Instead, its deviation from maximality is controlled\footnote{The special
case of eq.(\ref{mainconst}) for $\mathsf{\theta_\odot \approx 0}$, namely
$\mathsf{2 \sqrt{\Delta m^2_{LSND} \Delta m^2_\odot} \approx 
 \Delta m^2_{Atm}}$ was discovered in Ref. \cite{gaur}.} by the product of
the two factors in the RHS of (\ref{solpred}) and specifically by the
LSND mass scale. If we take $\mathsf{\Delta m^2_{LSND} \rightarrow \infty}$,
we are back to the three neutrino result of eq.(\ref{bound}). Moreover,
the range of the ratio $\mathsf{ (\Delta m^2_{Atm})^2 /
4 \Delta m^2_\odot \Delta m^2_{LSND} }$ with present experimental errors 
is such (\textbf{Fig 2}) that
both the small (SMA) and large mixing (LMA) MSW solutions are equally
allowed. This is quite unlike in the three generation case. 

\item 
The result, stated in eq.(\ref{solpred}), has been derived assuming the 
absence of any phases in $\mathcal{M}_\mathsf{\nu}^\mathsf{(4)}$ and hence
in $\mathsf{U}$. The presence of sizeable phases (\textit{i.e}
of a significant amount of CP-violation in the neutrino sector) can change
the result. 

\end{itemize}

\noindent
We infer that, in so far as the prediction for the solar neutrino mixing
angle is concerned, the radiative model \cite{gaur} with four neutrinos 
is on a significantly different footing as compared to the corresponding 
three generation version. Indeed, the former is better able
to tackle the  present data, especially from SNO. We discuss
in the next section some phenomenologically allowed textures of 
$\mathcal{M}_\mathsf{\nu}^\mathsf{(4)}$ from the current data. 

\section{Textures of the $\mathsf{4 \times 4}$ radiative model mass 
matrix }

Our derivation of the phenomenological relation eq.~(\ref{mainconst}) for
the mixing angle, $\mathsf{\theta_\odot}$ in Sec.~~\textbf{III} made use
of two inputs : (1) the $\mathsf{ 2 + 2}$ mass pattern implemented on the
neutrino mass matrix of eq. (\ref{fullmat}) with vanishing diagonal coefficients
and (2) the form of eq. (\ref{4gmns2})  for the $\mathsf{4 \times 4}$ 
unitary mixing matrix $\mathsf{U}$.  In this section, we estimate the orders of
magnitude of the nondiagonal entries of $\mathcal{M}_\mathsf{\nu}^\mathsf{(4)}$.
We do so by relating them to the mass scales $\mathsf{ (\Delta m^2_\odot)^{1/2},
~(\Delta m^2_{Atm})^{1/2}}$ and $\mathsf{(\Delta m^2_{LSND})^{1/2}}$  via the
elements of $\mathsf{U}$ and then constraining the latter from phenomenological
inputs. We also favor or disfavor certain mass textures of
$\mathcal{M}_\mathsf{\nu}^\mathsf{(4)}$. 

Let us parameterize the four real neutrino mass
eigenvalues, in a way consistent with eq. (\ref{4gconst1}), as
\begin{equation}
\label{param}
\mathsf{
m_{\nu_1} = m_1 \;,\;\; m_{\nu_2} = m_2 \;,\;\; m_{\nu_3} = M - m_1 \;, \;\; 
m_{\nu_4} = -M - m_2 .
}
\end{equation}

\noindent
In the \textbf{2 + 2} pattern, 
 $\mathsf{M}$ represents the mass scale of the heavier $\mathsf{(3,4)}$ pair,
being $\mathcal{O}(\mathsf{\sqrt{\Delta m^2_{LSND}}})$, 
while $\mathsf{|m_{1,2}|}$ stand for the mass magnitudes of the lighter 
$\mathsf{(1,2)}$ pair.  
Because of the constraint of eq. (\ref{mainconst}), we can treat $\mathsf{
(\Delta m^2_\odot)^{1/2}~ \sim | m_1^2 - m_2^2|}$ and
$\mathsf{(\Delta m^2_{LSND})^{1\over 2} \sim |M|}$ (see Fig. 1) 
as the two independent
controlling mass scales in $\mathcal{M}_\mathsf{\nu}^\mathsf{(4)}$. 
We have thus found in eq.(\ref{param}) a  parameterization 
that is convenient for the description of all neutrino oscillation data, obeys
the tracelessness condition eq.(\ref{4gconst1}) and 
follows the mass pattern depicted in Fig. 1. 

In order to estimate the orders of magnitude of those among
the nondiagonal elements of $\mathcal{M}_\mathsf{\nu}^\mathsf{(4)}$ 
which are nonvanishing, we resort to the equation
$\mathcal{M}_\mathsf{\nu}^\mathsf{(4)} = \mathsf{ U~ 
diag.\{ m_1, m_2, M-m_1, -M-m_2\}~ U^T}$. We have already put 
$\mathsf{U_{e3} \approx \epsilon,~ U_{e4} \approx \epsilon'}$ 
in eq.(\ref{4gmns2}) with
the expectation that $\mathsf{|\epsilon|,~|\epsilon'| ~\ler}~
 \mathcal{O}(\mathsf{10^{-2}})$  in view of eq.(\ref{bugeyconst}). We can 
also put 
$\mathsf{U_{\mu 1} \approx \delta,~U_{\mu 2} \approx \delta'}$ and expect
in the light of eq.(\ref{ccfrp}) that $\mathsf{|\delta|, ~|\delta'| ~\ler}~
\mathcal{O}(\mathsf{10^{-1}})$; indeed, the best fit requires them to be close
to zero, \textit{cf.} eq.(\ref{bestfit}). Furthermore, we can put 
$\mathsf{U_{\mu 3}^2 \approx U_{\mu 4}^2 \approx 1/2 }$ on account of the
observed maximal mixing in the atmospheric neutrino sector. The various 
entries of $\mathcal{M}_\mathsf{\nu}^\mathsf{(4)}$ can now be related to 
the mass eigenvalues and hence to $\mathsf{m_{1,2}}$ and $\mathsf{M}$ 
as follows:

\begin{eqnarray}
\mathsf{a} &\approx& \mathsf{m_1~(\delta~ U_{e1} - \epsilon~ U_{\mu 3} ) + 
m_2 ~(\delta'~ U_{e2} - \epsilon'~ U_{\mu 4} ) + M ~( \epsilon~ U_{\mu 3}  - 
 \epsilon'~U_{\mu 4}  )}, \\ 
\mathsf{b} &\approx& \mathsf{m_1~(U_{e1}~U_{\tau 1} - \epsilon~ U_{\tau 3} ) + 
m_2 ~(U_{e2}~U_{\tau 2} - \epsilon'~ U_{\tau 4} ) + M ~( \epsilon~ U_{\tau 3} 
- \epsilon'~U_{\tau 4})}, \\
\mathsf{c} &\approx& \mathsf{m_1~(\delta~U_{\tau 1} - U_{\mu 3}~ U_{\tau 3} ) + 
m_2 ~( \delta'~U_{\tau 2} - U_{\mu 4}~ U_{\tau 4} ) + M ~( U_{\mu 3}~ U_{\tau 3}
- U_{\mu 4} ~U_{\tau 4})}, \\
\mathsf{d} &\approx& \mathsf{m_1~(U_{e1}~U_{s1} - \epsilon~ U_{s 3} ) + 
m_2 ~( U_{e2}~U_{s 2} - \epsilon'~ U_{s 4} ) + M ~( \epsilon~ U_{s 3}
- \epsilon'~U_{s 4})}, \\
\mathsf{e} &\approx& \mathsf{m_1~(\delta~U_{s1} - U_{\mu 3}~ U_{s 3} ) + 
m_2 ~( \delta'~U_{s 2} - U_{\mu 4}~ U_{s 4} ) + M ~( U_{\mu 3}~ U_{s 3}
- U_{\mu 4} ~U_{s 4})}, \\
\mathsf{f} &=& \mathsf{m_1 ~( U_{s1} ~U_{\tau 1} - U_{s3}~ U_{\tau 3} ) + 
m_2~ ( U_{s2}~ U_{\tau 2} - U_{s4}~ U_{\tau 4 } ) + M ~(U_{s3}~ U_{\tau 3} 
- U_{s4}~ U_{\tau 4} ) }. 
\end{eqnarray}

\noindent
We see from the above that in the expressions for the $\mathsf{a,~b}$ and 
$\mathsf{d}$, the explicit occurrence of 
$\mathsf{\epsilon}$ and $\mathsf{\epsilon'}$ in the coefficient of $\mathsf{M}$
pulls it down significantly; thus they are much smaller in magnitude compared to
$\mathsf{M}$. In contrast, the coefficients of $\mathsf{M}$ in the expressions
for the entries $\mathsf{c,e}$ and $\mathsf{f}$ do not manifestly involve
such small parameters. This is a consequence of the fact that the
coefficients of $\mathsf{M}$ in 
all the entries depend only on the last two columns of the mixing matrix 
$\mathsf{U}$ eq.(\ref{4gmns2}). We see from eqs.(\ref{unit2}) -
(\ref{unit2a}), as also from eq.(\ref{param}) that, in the limit 
when $\mathsf{m_{1,2}~
\rightarrow ~0 ,~ U_{e3}^2 = U_{e4}^2,~U_{\mu 3}^2 = U_{\mu 4}^2 ,~ 
U_{\tau 3}^2 = U_{\tau 4}^2}$ and $\mathsf{U_{s3}^2 = U_{s4}^2}$. 
In such a limit, when $\mathsf{M}$ remains as  the only controlling mass 
scale,

\begin{eqnarray}
\mathsf{|\epsilon|}& \approx&  \mathsf{|\epsilon'|},\\ 
\label{casesa}
\mathsf{|U_{\mu 3}|}& \approx & \mathsf{|U_{\mu 4}|},\\ 
\label{casesb}
\mathsf{|U_{\tau 3}|}& \approx & \mathsf{|U_{\tau 4}|},\\ 
\label{casesc}
\mathsf{|U_{s 3}|}& \approx & \mathsf{|U_{s 4}|}. 
\end{eqnarray}

\noindent
We restate that we have assumed no phases to be present in our
mixing matrix. In the limit of neglecting terms $\mathcal{O}(\mathsf{\delta^2,
\delta^{'~2}})$, we then have\footnote{It follows from eq.(\ref{param}) that, 
when $\mathsf{m_{1,2}~ \rightarrow~ 0}$, $\mathsf{\nu_3}$ and $\mathsf{\nu_4}$
form a Dirac neutrino. So it is not surprising that we get maximal mixing
in the $\mathsf{3,~4}$ sector in this limit.} from the unitarity condition
$\mathsf{1 = U_{\mu 1}^2 + U_{\mu 2}^2 + U_{\mu 3}^2 + U_{\mu 4}^2 }$
and eq.(\ref{casesa}) that 
\begin{equation}
\label{atmmax}
\mathsf{
|U_{\mu 3}| = | U_{\mu 4}| = 1 /\sqrt{2}
}.
\end{equation}

In addition to satisfying the above conditions in the limit $\mathsf{m_{1,2} 
\rightarrow 0 }$, the parameters
$\mathsf{U_{s3}}$, $\mathsf{U_{s4}}$, $\mathsf{U_{\tau 3}}$, 
$\mathsf{U_{\tau 4} }$ also have to satisfy
the off-diagonal unitarity condition: 
\begin{equation}
\label{orthocond}
\mathsf{
U_{\mu 3}~ U_{\mu 4} + U_{\tau 3}~ U_{\tau 4} + U_{s3}~U_{s4} = ~0 ~+~ }
\mathcal{O}(\mathsf{\epsilon \epsilon'}). 
\end{equation}

\noindent
From eq.(\ref{atmmax}) we see that this would mean
\begin{equation}
\mathsf{
U_{s3}~U_{s4} + U_{\tau 3}~U_{\tau 4} \approx \pm { 1 \over \sqrt{2} }. 
}
\end{equation}
\noindent
Eqs. (\ref{casesa}), (\ref{casesb}) and (\ref{casesc}) admit four possibilities
 corresponding to the signs  each of the mixing matrix elements
can take. First, we  rewrite the condition (\ref{casesa}) in terms of two
possibilities as $\mathsf{U_{\mu 3}~ = ~U_{\mu 4}}$ or $\mathsf{ U_{\mu 3}
= - U_{\mu 4}}$. The remaining possibilities form sub-cases of these two cases.
Each of these (sub)cases is further constrained by the off-diagonal unitarity
condition of eq.(\ref{orthocond}). After some algebra, we realize that three
possible cases are consistent with the unitarity conditions. Each of them
leads to a different texture which we detail below. In all the following 
cases, we have  $\mathsf{a}$, $\mathsf{b}$, $\mathsf{c}$ $\sim \mathcal{O}(
\mathsf{m_{1,2}})$. 

\begin{itemize}
\item \textbf{Case A} Here, we estimate the following orders of magnitude
for the other entries of $\mathcal{M}_\mathsf{\nu}^\mathsf{(4)}$. 
\begin{equation} 
\mathsf{c ~ \approx ~2~ M~ U_{\mu 3}~U_{\tau 3}} \equiv \mathsf{M'};\;\;
\mathsf{e ~ \approx ~} \mathcal{O}(\mathsf{m_{1,2}});\;\;\;
\mathsf{f ~ \approx ~2~ M~ U_{s 3}~U_{\tau 3}} \equiv \mathsf{M''}. \nonumber
\end{equation}
This case comes either of the choices:  $\mathsf{U_{\mu 3}}$ 
$\mathsf{= U_{\mu 4}}$, $\mathsf{U_{\tau 3}}$ $\mathsf{=- U_{\tau 4}}$,
$\mathsf{ U_{s3}}$ $\mathsf{=~U_{s4}}$ or   
$\mathsf{U_{\mu 3}}$ $\mathsf{=~-U_{\mu 4}}$, $\mathsf{U_{\tau 3}}$ 
$\mathsf{=U_{\tau 4}}$, $\mathsf{U_{s3}}$ $\mathsf{=-U_{s4}}$. 
The mass texture would be of the form : 

\be
\mathcal{M}_{\mathsf{\nu}}^\mathsf{(4)}  \sim 
\left( 
\begin{array}{cccc}
\mathsf{0}&\mathsf{a}&\mathsf{b}&\mathsf{d} \\
\mathsf{a}&\mathsf{0}&\mathsf{M'}&\mathsf{e} \\
\mathsf{b}&\mathsf{M'}&\mathsf{0}&\mathsf{M''}\\
\mathsf{d}&\mathsf{e}&\mathsf{M''}&\mathsf{0} 
\end{array}
\right), 
\ee 
\noindent
The active sterile admixture $\mathsf{A}$ is now given by
\be
\mathsf{
A \approx 1 - {f^2 \over c^2} 
}.
\ee 
We see that\footnote{The unitarity condition $\mathsf{1}$ $=$ 
$\mathsf{U_{s1}^2}$ $+$ $\mathsf{U_{s2}^2}$ $+$ $\mathsf{U_{s3}^2}$ $+$
$\mathsf{U_{s4}^2}$ and eq. (\ref{casesc}) imply that 
 $\mathsf{f^2/c^2}$ $\approx$ $\mathsf{|U_{s3}| \ler 1/2}$.}
$\mathsf{2~U_{s3}^2}$ $\mathsf{\le 1}$. 
Thus we can parameterize $\mathsf{f/c}$ as $\mathsf{cos~\beta}$ in which
case $\mathsf{A = sin^2 ~\beta}$. 

\item \textbf{Case B} Here the various magnitudes that $\mathsf{c,~e,~f}$ 
can take are given by
\begin{equation} 
\mathsf{c ~ \approx }~ \mathcal{O}(\mathsf{m_{1,2}});\;\;\;
\mathsf{e ~ \approx ~2~ M~ U_{\mu 3}~U_{s 3}} \equiv \mathsf{M'''};\;\;
\mathsf{f ~ \approx ~2~ M~ U_{s 3}~U_{\tau 3}} \equiv \mathsf{M''}. \nonumber
\end{equation}
This situation can arise with either of the choices : 
$\mathsf{U_{\mu 3}}$ $=$ $\mathsf{U_{\mu 4}}$, $\mathsf{U_{\tau 3}}$ $=$ 
$\mathsf{U_{\tau 4}}$, $\mathsf{U_{s3}}$ $=$ $\mathsf{-U_{s4}}$ or 
$\mathsf{U_{\mu 3}}$ $=$ $\mathsf{-U_{\mu 4}}$, $\mathsf{U_{\tau 3}}$ 
$=$ $\mathsf{-U_{\tau 4}}$, $\mathsf{U_{s3}}$ $=$ $\mathsf{U_{s4}}$. 
The mass matrix would now take the form: 

\be
\mathcal{M}_{\mathsf{\nu}}^\mathsf{(4)} \sim
\left( 
\begin{array}{cccc}
\mathsf{0}&\mathsf{a}&\mathsf{b}&\mathsf{d} \\
\mathsf{a}&\mathsf{0}&\mathsf{c}&\mathsf{M'''} \\
\mathsf{b}&\mathsf{c}&\mathsf{0}&\mathsf{M''}\\
\mathsf{d}&\mathsf{M'''}&\mathsf{M''}&\mathsf{0} 
\end{array}
\right). 
\ee 
The active sterile admixture in this case is given by
\be
\mathsf{
A \approx 1 - {f^2 \over e^2}} .
\ee 
Once again, $\mathsf{f^2/e^2}$ $=$ $\mathsf{2~U_{s3}^2}$ and can be
parameterized as $\mathsf{cos^2 \beta}$  so that $\mathsf{A}$ $=$
$\mathsf{sin^2~\beta}$. 

\item \textbf{Case C}
Here we have
\begin{equation} 
\mathsf{c ~ \approx ~2~ M~ U_{\mu 3}~U_{\tau 3}} \equiv \mathsf{M'};\;\;
\mathsf{e ~ \approx ~2~ M~ U_{\mu 3}~U_{s 3}} \equiv \mathsf{M'''};\;\;
\mathsf{f ~ \approx~ } \mathcal{O}(\mathsf{m_{1,2}}),
\end{equation}
arising from either of the choices : $\mathsf{U_{\mu 3}}$ $=$ 
$\mathsf{U_{\mu 4}}$, $\mathsf{U_{\tau 3}}$ $=$ $\mathsf{U_{\tau 4}}$,
$\mathsf{U_{s3}}$ $=$ $\mathsf{U_{s4}}$ or $\mathsf{U_{\mu 3}}$ $=$ 
$\mathsf{-U_{\mu 4}}$, $\mathsf{U_{tau 3}}$ $=$ $\mathsf{-U_{\tau 4}}$, 
$\mathsf{U_{s3}}$ $=$ $\mathsf{-U_{s4}}$. 
 In this case the mass matrix has the following texture
\be
\label{texturec}
\mathcal{M}_{\mathsf{\nu}}^\mathsf{(4)} =  
\left( 
\begin{array}{cccc}
\mathsf{0}&\mathsf{a}&\mathsf{b}&\mathsf{d} \\
\mathsf{a}&\mathsf{0}&\mathsf{M'}&\mathsf{M'''} \\
\mathsf{b}&\mathsf{M'}&\mathsf{0}&\mathsf{f}\\
\mathsf{d}&\mathsf{M'''}&\mathsf{f}&\mathsf{0} 
\end{array}
\right). 
\ee 
The active sterile admixture is now given by
\be
\mathsf{
A \approx {c^2 \over c^2 ~+~ e^2} 
}.
\ee 
The form $\mathsf{A}$ $=$ $\mathsf{sin^2~\beta}$ can now be achieved by 
parameterizing $\mathsf{e/c}$ $=$ $\mathsf{U_{\tau 3}/U_{s3}}$ as 
$\mathsf{tan~ \beta}$.
We note that this texture has been recently been studied in 
Ref. \cite{asjsruba} on the basis of two pseudo-Dirac neutrinos and an
approximate global $\mathsf{L_e ~+~L_\mu~-~L_\tau~-~L_s}$ symmetry. 
A slight variation of 
eq. (\ref{texturec}) with $\mathsf{f}$ $=$ $\mathsf{0}$ and a nonzero fourth
diagonal element has been proposed in Ref.\cite{babumohap} on the basis of
a paired pseudo-Dirac structure and the same global symmetry, but not 
within a Zee-type radiative model. The textures of our cases \textbf{A} and
\textbf{B} strongly violate the global $\mathsf{L_e + L_\mu - L_\tau - L_s}$
symmetry.  
\end{itemize}

All the above textures give rise to the \textbf{2 + 2} pattern along with
an admixture which in general is between zero or one. The choice 
$\mathsf{U_{s3}}$ $=$ $\mathsf{0}$ or $\mathsf{U_{s3}^2}$ $=$ $\mathsf{1/2}$
will yield the extreme results $\mathsf{A}$ $=$ $\mathsf{0}$ or $\mathsf{1}$
respectively, but such a choice is not favored by the experimental data
\cite{concha}. In all the above three cases
the LSND scale appears in \textit{two} pairs of elements in 
$\mathcal{M}_\mathsf{\nu}^\mathsf{(4)}$ if $\mathsf{\beta}$ is nonzero.
This is not surprising since, for a nonvanishing admixture $\mathsf{A}$, the
sterile species $\mathsf{\nu_s}$ and the tau neutrino $\mathsf{\nu_\tau}$
appear in both the lighter $\mathsf{(1,2)}$ and the heavier $\mathsf{(3.4)}$
pair of physical states. 

\section{conclusions}
In this paper we have studied the constraints on the neutrino mass matrix
$\mathcal{M}_{\nu}$ for a radiative model \cite{gaur} of three active
and one sterile neutrino species from all the current data on neutrino
oscillations. We have established a method to show how the solar neutrino
mixing angle is related to the observed squared mass differences. This leads,
in the three neutrino case to a near-maximal $\mathsf{\theta_\odot}$,
not in favor with the latest data. In the four neutrino case, however,
we derive the result 
$\mathsf{ sin^2 2 \theta_\odot \approx
 1 - [(\Delta m^2_{Atm})^2/(4~ \Delta m^2_{LSND} \Delta m^2_\odot)]^2 }$
which allows a phenomenologically  acceptable value for $\mathsf{\theta_\odot}$
within the allowed ranges of the three squared mass differences. The above
result is compatible with the present solar neutrino data when MSW solutions
with either large or small mixing are considered. However, it is difficult
to reconcile LOW-QVO solutions within these models. 

We have shown then that the radiative model with four neutrinos 
offers several phenomenologically viable textures of 
$\mathcal{M}_\mathsf{\nu}^\mathsf{(4)}$. This is quite 
unlike in the three neutrino case where maximal mixing in the 
atmospheric sector forces the  experimentally disfavored
bimaximal texture.

\vskip 0.5cm
\noindent
\textbf{Acknowledgments:}
We wish to thank Anjan Joshipura, Uma Sankar and Mohan Narayan for fruitful
discussions. 

\appendix
\section*{Appendix}
Here we present the neutrino mass solutions of 
eqs.(\ref{4gsimul1}) - (\ref{4gsimul3}). There  are four sets of solutions.
These can be conveniently expressed in terms of numerators 
$\mathsf{N^\pm_{1,2,3}}$ and common denominators $\mathsf{D^\pm}$ defined
as follows.
\begin{table}[t]
\label{Table 2}
\begin{tabular}{|c|c|c|c|}
\hline
\textbf{Set 1}& \textbf{Set 2}& \textbf{Set 3}& \textbf{Set 4} \\[10pt]
\hline
$\mathsf{m_{\nu_1}~=~{N_1^+/D^+}}$ &
$\mathsf{m_{\nu_1}~=~-{N_1^+/D^+}}$ &
$\mathsf{m_{\nu_1}~=~{N_1^-/D^-}}$ &
$\mathsf{m_{\nu_1}~=~-{N_1^-/D^-}}$ \\[10pt]
\hline
$\mathsf{m_{\nu_2}~=~{N_2^+/D^+}}$ &
$\mathsf{m_{\nu_2}~=~-{N_2^+/D^+}}$ &
$\mathsf{m_{\nu_2}~=~{N_2^-/D^-}}$ &
$\mathsf{m_{\nu_2}~=~-{N_2^-/D^-}}$ \\[10pt]
\hline
$\mathsf{m_{\nu_3}~=~{N_3^+/D^+}}$ &
$\mathsf{m_{\nu_3}~=~-{N_3^+/D^+}}$ &
$\mathsf{m_{\nu_3}~=~{N_3^-/D^-}}$ &
$\mathsf{m_{\nu_3}~=~-{N_3^-/D^-}}$ \\[10pt]
\hline
\end{tabular}
\caption{Neutrino mass solutions} 
\end{table}

\begin{eqnarray}
\mathsf{N^+_1} &\equiv&  \mathsf{-~4~ \Delta m^2_{LSND}~ \Delta m^2_\odot 
+  (\Delta m^2_{Atm})^2 - 2~ \Delta m^2_\odot~\Delta m^2_{Atm} 
-~3 ~ (\Delta m^2_\odot)^2 } ,\nonumber \\[10pt]
\mathsf{N^+_2} &\equiv&  \mathsf{4~ \Delta m^2_{LSND}~ \Delta m^2_\odot 
+  (\Delta m^2_{Atm})^2 + 2~ \Delta m^2_\odot~\Delta m^2_{Atm} 
+~ (\Delta m^2_\odot)^2 }, \nonumber \\[10pt]
\mathsf{N^+_3} &\equiv&  \mathsf{4~ \Delta m^2_{LSND}~ \Delta m^2_{Atm}
+  (\Delta m^2_{Atm})^2 + 2~ \Delta m^2_\odot~\Delta m^2_{Atm} 
+~ (\Delta m^2_\odot)^2 }, \nonumber \\[10pt]
\mathsf{D^+}&\equiv& \mathsf{ 2 \sqrt{2}~\left[ \left\{ (\Delta m^2_{Atm})^2 
- (\Delta m^2_\odot)^2 
\right\} \left( \Delta m^2_{Atm} + 2 ~\Delta m^2_{LSND} + \Delta m^2_\odot 
\right) \right]^{1 \over 2} }, \nonumber
\end{eqnarray}
and
\begin{eqnarray}
\mathsf{N^-_1} &\equiv&  \mathsf{~4~ \Delta m^2_{LSND}~ \Delta m^2_\odot 
+  (\Delta m^2_{Atm})^2 + 2~ \Delta m^2_\odot~\Delta m^2_{Atm} 
-~3 ~ (\Delta m^2_\odot)^2 }, \nonumber \\[10pt]
\mathsf{N^-_2} &\equiv&  \mathsf{-~ 4~ \Delta m^2_{LSND}~ \Delta m^2_\odot 
+  (\Delta m^2_{Atm})^2 - 2~ \Delta m^2_\odot~\Delta m^2_{Atm} 
+~ (\Delta m^2_\odot)^2 }, \nonumber \\[10pt]
\mathsf{N^-_3} &\equiv&  \mathsf{4~ \Delta m^2_{LSND}~ \Delta m^2_{Atm}
+  (\Delta m^2_{Atm})^2 - 2~ \Delta m^2_\odot~\Delta m^2_{Atm} 
+~ (\Delta m^2_\odot)^2 }, \nonumber \\[10pt]
\mathsf{D^-}&\equiv& \mathsf{ 2 \sqrt{2}~\left[ \left\{ (\Delta m^2_{Atm})^2 
- (\Delta m^2_\odot)^2 
\right\} \left( \Delta m^2_{Atm} + 2 ~\Delta m^2_{LSND} - \Delta m^2_\odot 
\right) \right]^{1 \over 2} }. \nonumber
\end{eqnarray}

\noindent
Note that $\mathsf{N^-_i}$ $(\mathsf{i = 1, 2, 3})$ and $\mathsf{D^-}$ 
can always be obtained from the corresponding 
$\mathsf{N^+_i}$ and $\mathsf{D^+}$ by putting 
$\mathsf{\Delta m^2_\odot}$ $\rightarrow$
$- \mathsf{\Delta m^2_\odot}$.  
The four sets of neutrino mass solutions  with
$\mathsf{m_{\nu_4}}$ always being 
$\mathsf{- m_{\nu_1} - m_{\nu_2} - m_{\nu_3}}$, are shown in  table I.

\end{document}